# Ab-initio determination of exchange integrals and Néel temperature in the chain cuprates

A. B. van Oosten[a] and Frédéric Mila[b]

[a]Max Planck Institut für Physik komplexer Systeme, Nöthnitzerstrasse 38, 01187 Dresden, Germany

[b]Laboratoire de Physique Quantique, Université Paul Sabatier, 31062 Toulouse France

We report ab initio quantum chemical cluster calculations of the chain ($J_a$) and the largest interchain ($J_b$) Heisenberg exchange of the chain cuprates $Ca_2CuO_3$ and $Sr_2CuO_3$. We find that $J_a$ is comparable to the in-plane J in layered cuprates and $J_a/J_b \approx 250-400$. Using recent theory we obtain close agreement with experiment for the staggered magnetic moments and critical temperatures. This implies that $T_N$ does not depend on the third parameter, $J_c \ll |J_b|$, and cannot be calculated using spin-wave theory. We propose an explanation of this fact in terms of a 1D-->2D cross-over.

The one-dimensional (1D) spin 1/2 Heisenberg antiferromagnet is a nontrivial, but nonetheless integrable [1] quantum mechanical model with vanishing long-range magnetic order in the ground state. In real systems containing antiferromagnetic (AF) chains, interchain coupling often hinders the study of the 1D behaviour. The linear cuprates $Ca_2CuO_3$ and $Sr_2CuO_3$ [2] present a unique opportunity to compare exact results to experiment because of their exceptionally small ratio of inter- to intrachain coupling. The crystal structure consists of two interpenetrating, magnetically decoupled sublattices of infinite chains of S=1/2 $[CuO_3]^{-4}$ units separated by $Ca(Sr)^{+2}$ ions [3,4]. A single sublattice can be described by the Heisenberg hamiltonian where the summation runs over lattice vectors (ijk) and the unit cell convention is defined in Fig. 1. The magnetic susceptibility is entirely determined by the isolated chains, i.e. by the intrachain AF coupling $J_a$. Two independent experiments yield incompatible values $-J_a \approx 135\text{-}160$ meV [3,5,6] and $J_a = -190$ meV [7]. An even larger value of $J_a = -260$ meV was estimated from the analysis of the mid-gap infrared absorption spectrum. For comparison, in the planar cuprates the in-plane coupling varies between -100 and -140 meV[8,9]. A fascinating phenomenon is the transition to AF order at exceptionally low Néel temperature[10-12]. A recent theory developed by Schulz [13] predicts that the Néel temperature varies linearly with $J_\perp$ for isotropic coupling of the chains and that the square of the magnetic moment $m_\pi$ is proportional to the Néel temperature $T_N$. This theory was applied to the problem of a two dimensional array of chains at zero temperature with the conclusion that such a system exhibits long range-order[14]. Recent muon spin rotation and neutron scattering experiments on $Ca_2CuO_3$ and $Sr_2CuO_3$ yield $T_N \propto m_\pi^2$, approximately, and $J_\perp$ of the order of 1 meV. From a LDA band structure calculation $J_b$ has been estimated to be 3.6 meV and 0.8 meV for $Ca_2CuO_3$ and $Sr_2CuO_3$, respectively [15]. However, Schulz's theory should not apply to the present case. It assumes isotropic interchain coupling, whereas in $Ca_2CuO_3$ and $Sr_2CuO_3$ the ferromagnetic [11] coupling along the c-direction should be extremely small, i.e. $J_c << |J_b|$. Therefore, Schulz's theory predicts that $T_N$ and $m_\pi$ depend only on the average of $J_b$ and $J_c$. This is in sharp contrast to the fact that, in the limit $J_c \to 0$, $m_\pi \to 0$ at finite T according to the Hohenberg-Mermin-Wagner (HMW) theorem [16,17].

In this Letter we first determine the value of $J_a$ with an approach that is entirely independent from experiment and, for the quasi-2D cuprates, rivals with experiment in accuracy [8,9]. We then compute $J_b$ and show that Schulz's theory accurately predicts $m_\pi$ as well as $T_N$, even though $J_c << J_b$. We then explain this unexpected result in a manner consistent with the HMW theorem.

The computational approach for the intrachain coupling $J_a$ follows that of Refs. [8,9]. It is an accurately balanced calculation of singlet-triplet splitting of a cluster containing calculate the (MC)SCF singlet and triplet ground states of the planar, $D_{2h}$ symmetric, $Cu_2O_7$ cluster that occurs in the chains. As demonstrated in Refs.[8,9], it is necessary and adequate to introduce local exchange and correlation effects on the bridging oxygen atom through subsequent admixture of excited (MC)SCF states, that differ from the MCSCF ground states by an O -> Cu electron excitation. We follow the non-orthogonal configuration interaction (NOCI) approach [19], which involves the computation of Hamilton and overlap matrix elements between determinants constructed from non-orthogonal orbital sets [20]. The NOCI approach leads to short, physically transparent wavefunctions and is size consistent.

For the calculation of the interchain coupling $J_b$, we perform all-electron (MC)SCF calculations of the singlet and triplet ground states of $D_{2h}$ symmetric $Cu_2O_8$ and $Cu_2O_8M_{16}$ (M=Ca,Sr) clusters (Fig. 2) consisting of two $CuO_4$ units that are translated with respect to each other by one unit vector along the b-axis, so that they are located in next-nearest neighbour chains. Since there are no ligands, the exchange originates simply from the overlap between the Cu d-orbitals and $J_b$ can be found directly from the (MC)SCF ground state singlet-triplet splitting. For a study of the effect of the counterions, calculations were also performed on $Cu_2O_8M_{16}$ (M=Ca,Sr).

The clusters are embedded in a point charge environment with modified potentials at the nearest neighbour positions to the cluster. The complete specification of the background potential is available from the authors on request. For Cu and O the same contracted Gaussian basis sets are used as in ref. [9], namely a 14s11p6d -> 8s6p3d basis set for Cu and a 9s6p -> 3s3p one for O. It was found that $J_b$ is sensitive to extension of the Cu basis with diffuse d-functions, as will be discussed below in detail. For Ca we employed a 14s8p -> 5s2p basis set of double zeta quality [21] and a 12s6p -> 4s2p minimal basis set [22] and for Sr a (15s,9p,3d) -> (5s,3p,1d) minimal basis set [22].

Following Ref. [8,9], a first approximation to the intrachain coupling is obtained from a complete active space self consistent field (CASSCF) calculation of the singlet-triplet splitting of a $D_{2h}$ symmetric $[Cu_2O_7]^{-10}$ cluster model. The active orbitals (whose occupation are allowed to vary) are even (g) and odd (u) linear combinations, $d_g$ and $d_u$,

of Wannier-like orbitals, $d_1$ and $d_2$, localised at each Cu. The triplet wavefunction can be written as

$$\Psi_t = |\sigma\bar{\sigma}d_g\bar{d}_u| = |\sigma\bar{\sigma}d_1 d_2| . \qquad (2)$$

Here $\sigma$ denotes the O($2p_\sigma$) orbital at the bridging oxygen, which has the same symmetry as $d_u$. The other closed shell orbitals are suppressed in the notation. The singlet corresponding to Eq.(2) can be written as

$$\Psi_s = \frac{(1+S)|\sigma\bar{\sigma}d_g\bar{d}_g| - (1-S)|\sigma\bar{\sigma}d_u\bar{d}_u|}{\sqrt{(2+2S^2)}} . \qquad (3)$$

In Eq. (3) occurs the overlap $S = \langle d_1 | d_2 \rangle$ as an additional variational parameter. As usual, the triplet and singlet SCF ground states are very well characterised by $Cu^{+2}$ ($3d^9$) and $O^{-2}$ ($2p^6$) and the Cu holes have almost pure $3d(x^2-y^2)$ character in a Mulliken orbital population. We find $S \approx 0.04$ and $E_t - E_s = 20$-$30$ meV. As before we admix to $\Psi_t$ and $\Psi_s$ relaxed charge transfer (CT) excitations of the form

$$\Psi_t^* = |d_u\bar{d}_u d_g\sigma| , \qquad (4)$$

$$\Psi_s^* = \frac{|d_g\bar{d}_g d'_u\bar{\sigma}'| - |d_g\bar{d}_g\bar{d}'_u\sigma'|}{\sqrt{(2+2S^{*2})}} . \qquad (5)$$

In the excited singlet state $\Psi_s^*$ (5) $d'_u$ and $\sigma'$ are two non-orthogonal linear combinations of $\sigma$ and $d_u$. As before $d'_u$ and $\sigma'$ are strongly overlapping, $S^* \approx 0.55$, and $\Psi_s^*$ describes a resonating covalent Cu-O bond. The subsequent NOCI calculation involves a 2x2 non-orthogonal diagonalisation for the triplet and a 5x5 one for the singlet, since in the latter S and the coefficients defining $d'_u$ and $\sigma'$, namely $S^*$ and a $d_u/\sigma$ mixing angle, are reoptimised. The resulting ground and excited states are separated by about 10 eV. For both $Ca_2CuO_3$ and $Sr_2CuO_3$ a value of $J_a = -119$ meV is obtained, nearest Ca neighbours are included, as all electron atoms, the value for $Ca_2CuO_3$ increases to 136.5 meV (Table I). From this it is plausible that also the $Sr_2CuO_3$ value will be between –119 and –136 meV. Our results corroborate analyses [5,6] of the susceptibility data [3] and are evidence against much larger values such as $J_a$=–190 meV [7] and $J_a$=–260 meV [23].

The interchain exchange $J_b$ equals the singlet-triplet splitting of the $Cu_2O_8$ based clusters shown in Fig. 2. There are no bridging O so that the (MC)SCF calculation following Eqs. (1) and (2) is sufficient. In this calculation the orbital $d_u$ transforms as $b_{1u}$. Since the Cu-Cu distance of 3.278 Å ($Ca_2CuO_3$) and 3.494 Å ($Sr_2CuO_3$) is rather larger than a Cu-Cu inter-atomic distance, we study the effect on $J_b$ of basis set extension with diffuse Cu d- and O p-functions for $[Cu_2O_8]^{-12}$ clusters. The effect of diffuse d-functions saturates with addition of $\alpha$=0.05, 0.016 and 0.005 bohr$^{-2}$ functions, while an extra $\alpha$=0.04 bohr$^{-2}$ O p-function has negligible effect (Table II). The diffuse functions contribute little to the total energy. The interchain coupling saturates towards values of $J_b$=0.33 meV ($Ca_2CuO_3$) and $J_b$=0.23 meV ($Sr_2CuO_3$). The overlaps of the d-orbitals are ~ 4 ‰. The calculation was repeated for $Cu_2O_8(Ca,Sr)_{16}$ clusters shown in Fig. 2, using the extended Cu basis. The effect of the counterions is an increase of $|J_b|$ from 0.33 meV to 0.52 meV for $Ca_2CuO_3$ and from 0.23 meV to 0.28 meV for $Sr_2CuO_3$. The same results were obtained for the double zeta and the minimal Ca basis set. It is a reasonable assumption that further extension of the cluster will have negligible effect.

From the fact that only the diffuse d-functions produce the antiferromagnetic interaction it is clear that $J_b$ originates from overlap of the diffuse tails of the Cu $3d(x^2-y^2)$ orbitals. In the language of the Hubbard model, one may describe this by a model system with a ground and a charge transfer excited state, connected by a matrix element 2t and separated in energy by U-V in the limit t –> 0 [15]. Here U is the on-site repulsion in the d-orbital and V the Coulomb attraction between the electron and the hole it leaves behind. The ground state can be identified with the singlet $(|d_1\bar{d}_2|-|\bar{d}_1d_2|)/\sqrt{2}$ and the excited state with $(|d_1\bar{d}_1|+|d_2\bar{d}_2|)/\sqrt{2}$. Solving the 2×2 Hamiltonian and comparing to Eq. (3) one obtains the admixture of the charge transfer state as S=–2t/(U-V) and the energy lowering the relaxed non-orthogonal singlet and the overlap S, one can therefore determine the parameters t and U-V (Table IV). As one would expect, t decreases and U-V increases with increasing Cu-Cu distance. The increase of U-V of 0.3-0.4 eV can be ascribed to a decrease of V by 0.3 eV, as expected for two pointcharges. This suggests that V≈4 eV and U≥23 eV. This is larger than the atomic estimate $E(Cu^{3+})+E(Cu^{+})-2E(Cu^{++})\approx 17$ eV. Note, however, that U relates to the atomic 3d orbital, while only the diffuse tail of the $3d(x^2-y^2)$ orbital is involved in the hopping. The values for U and t are larger than values adopted in an LDA band structure calculation [15] values of 3-3.5 eV. Nonetheless, the large value reported here makes sense in view of the good experimental agreement obtained for $J_b$.

In Refs. [13,14] a theory of weakly coupled AF chains is developed. It is based on the response of a chain to a staggered magnetic field, $\chi_{st}^{1D}(T)=0.32/T \ln^{1/2}(5.8J/T)$ [24]. The interchain coupling is treated in the random phase approximation (RPA). In the present, anisotropic case the resulting expressions for $m_\pi$ and $T_N$ are

$$m_\pi = 1.017 \sqrt{|J_\perp/J|} \quad (7)$$

and

$$T_N = 1.28 |J_\perp| \ln^{1/2}\left|\frac{5.8J}{T_N}\right|. \quad (8)$$

Eq. (7) should be applicable to the present anisotropic case by taking $|J_\perp|=(|J_b|+|J_c|)/2=|J_b|/2$, as we can put $J_c=0$ As can be seen from Table III, the resulting values of $m_\pi$ are in excellent agreement with experiment [12]. For the reasons stated above, Eq. (8) should not be valid in the present case. Nonetheless, the approximate experimental relationship $T_N \propto m_\pi^2$ encouraged us to apply Eq. (8) anyway. The results are in surprisingly accurate agreement with experiment, as can be seen from Table III. The situation is graphically displayed in Fig. 3.

We suggest an explanation of unexpected validity of Eq. (8). Roughly speaking [25] the 3D staggered susceptibility in the RPA can be written

$$\chi_{st}^{3D}(T) = \frac{\chi_{st}^{1D}(T)}{1 - (z_b|J_b|+z_c|J_c|)\,\chi_{st}^{1D}(T)},$$

where $z_i$ is the number of neighbours in the i-direction. This is equivalent to the set of expressions

$$\chi_{st}^{3D}(T) = \frac{\chi_{st}^{2D}(T)}{1 - z_c|J_c|\,\chi_{st}^{2D}(T)} \qquad (9)$$

$$\chi_{st}^{2D}(T) = \frac{\chi_{st}^{1D}(T)}{1 - z_b|J_b|\,\chi_{st}^{1D}(T)}. \qquad (10)$$

As argued above, $\chi_{st}^{2D}(T)$ cannot diverge at finite T, but is limited by fluctuations not present in the RPA treatment, so that Eq. (10) is not correct. For the case $J_a=J_b$ it is known that $\chi_{st}^{2D}(T)$ diverges as $T^{-1}\exp(AJ/T)$, where A is a numerical constant [26] and similar behaviour is plausible in the anisotropic case. Instead of Eq. (10) we propose the behaviour that is graphically displayed in Fig. 4. If $J_c$ is identically zero, the staggered susceptibility changes from $\chi_{st}^{1D}$ to the unknown form $\chi_{st}^{2D}$ at $T^*$, where $T^*$ is still determined by the condition $1=z_b|J_b|\,\chi_{st}^{1D}(T^*)$. For non-vanishing $J_c$, besides the cross-over at $T^*$, a Néel transition occurs at $T_N<T^*$, where $T_N$ is determined by the condition $1=z_c|J_c|\,\chi_{st}^{2D}(T_N)$ whereby $T_N$ is guaranteed to vanish with $J_c$ as required by HMW. This is true as long as $|J_c|$ is smaller than a critical value $J_c^*$, determined by the condition that $T_N=T^*$. From the fact that $\chi_{st}^{2D}(T)>>\chi_{st}^{1D}(T)$ unless $T>>T^*$, it is clear that $J_c^*<<J_b$. When $|J_c|>J_c^*$ one has $T_N>T^*$ but, since $\chi_{st}^{2D}(T)$ rapidly reduces to $\chi_{st}^{1D}(T)$ for $T>T^*$, $T_N$ will remain very close to $T^*$. Thus in this case the Néel transition takes places very near to $T^*$ as given by Schulz's theory. In Fig. 5 an example of Schulz's prediction for $T^*$ and the linear spin wave (LSW) estimate for $T_N$ are plotted against $J_c$. For $|J_c|<J_c^*$ the LSW estimate is qualitatively correct, while for $|J_c|>J_c^*$ it fails to reproduce the pinning of $T_N$ to $T^*$. In this regime Schulz's prediction of $T^*\cong T_N$ is quantitatively correct. Monte Carlo calculations on a plane system of 10 chains each containing 100 spins and with a

Finally, it should be possible to experimentally observe the change in critical behaviour when the dimensional cross-over transforms into the Néel transition. Moreover, the behaviour discussed here should be universal and should also occur in systems other than magnetic ones that exhibit strong anisotropy and a transition from 1D to 3D correlation.

In summary, the intra- and interchain magnetic coupling, $J_a$ and $J_b$, in $Ca_2CuO_3$ and $Sr_2CuO_3$ were computed. We obtain values of $J_a$=-119 meV for the intrachain coupling in both $Ca_2CuO_3$ and $Sr_2CuO_3$, respectively, with $Cu_2O_7$ clusters. This value is close to the value previously obtained for $La_2CuO_4$. For the $Cu_2O_7M_{16}$ (M=Ca,Sr) clusters $J_a$ is enhanced to -136 meV, which agrees with several analyses of the magnetic susceptibility of $Sr_2CuO_3$. For the interchain coupling we obtain $J_b$=-0.52 meV and $J_b$=-0.28 meV, respectively. With these values Schulz-'s theory [13,14] predicts magnetic moments and Néel temperatures in close agreement with experimental data. Thus the transition temperature is determined by Heisenberg coupling in 2D only. We propose an, in principle, universal mechanism which pins the Néel temperature is to the dimensional (1D->2D) cross-over temperature $T^*$, which is determined only by $J_a$ and $J_b$, if the third coupling $|J_c|$ exceeds a certain critical value $J_c^* \ll |J_b|$. We suggest that an experimental study of the critical behaviour of $Ca_2CuO_3$ and $Sr_2CuO_3$ should be rewarding.

We thank Stephan-Ludwig Drechsler, Walter Stephan and Karlo Penc for stimulating discussions.

# Tables

|  | a<br>meV | b<br>meV | c<br>meV |
|---|---|---|---|
| $Sr_2CuO_3$ ($Cu_2O_7$) | −32.8 | −118.7 | −100..−260[1] |
| $Ca_2CuO_3$ ($Cu_2O_7$) | −36.5 | −118.8 | — |
| $Ca_2CuO_3$ ($Cu_2O_7Ca_{16}$) | −48.1 | −136.5 | — |

Table I. Calculated intrachain coupling $J_a$ of the linear cuprates $Sr_2CuO_3$ and $Ca_2CuO_3$: a) Triplet-singlet splitting of Eqs. (2,3); b) Non-orthogonal CI involving (2,3) and (4,5); d) Experiment. [1]Refs. [3,5-7,23].

| added diffuse exponent | | $J_b$ (meV) | |
|---|---|---|---|
| O 2p | Cu 3d | $Sr_2CuO_3$ | $Ca_2CuO_3$ |
| — | — | −0.167 | −0.209 |
| — | 0.05 | −0.176 | −0.252 |
| — | 0.05, 0.016 | −0.226 | −0.321 |
| — | 0.05, 0.016, 0.005 | −0.228 | −0.330 |
| 0.04 | — | −0.166 | −0.213 |

Table II. Calculated interchain coupling $J_b$ of $Sr_2CuO_3$ and $Ca_2CuO_3$ from $Cu_2O_8$ clusters, for various basis sets.

|                                   | $J_b$ (meV) | magnetic moment ($\mu_B$) |            | Néel temperature (K) |            |
| --------------------------------- | ----------- | ------------------------- | ---------- | -------------------- | ---------- |
|                                   |             | theory[a]                 | experiment[b] | theory[a]         | experiment[b] |
| $Sr_2CuO_3$                       | –0.276      | 0.065                     | 0.06(1)    | 5.58                 | 5.41(1)    |
| $Ca_2CuO_3$                       | –0.522      | 0.089                     | 0.09(1)    | 10.12                | 10         |

Table III. Calculated interchain coupling $J_b$ along with ensuing theoretical values for staggered magnetic moment and Néel temperature in $Sr_2CuO_3$ and $Ca_2CuO_3$. $J_b$ is calculated using $Cu_2O_8M_{16}$ (M=Sr,Ca) clusters. Theoretical values were obtained from the theory described in Refs. [13,14]. Experimental values were taken from Ref. [12].

|  | 2t | U-V |
|---|---|---|
|  | meV | eV |
| $Sr_2CuO_3$ ($Cu_2O_8$) | –75 | 19.9 |
| $Ca_2CuO_3$ ($Cu_2O_8$) | –87 | 19.6 |
| $Sr_2CuO_3$ ($Cu_2O_8Sr_{16}$) | –83 | 19.2 |
| $Ca_2CuO_3$ ($Cu_2O_7Ca_{16}$) | –109 | 18.8 |

Table IV. Calculated hopping parameter t and on-site repulsion minus Coulomb attraction U-V of the linear cuprates $Sr_2CuO_3$ and $Ca_2CuO_3$.

## Figure captions

FIG. 1. The $M_2CuO_3$ (M=Sr,Ca) unit cell. $CuO_4$ 'plaquettes' are comnpleted with O outside the unit cell.

FIG. 2. The $Cu_2O_8M_{16}$ (M=Sr,Ca) cluster.

FIG. 3. Staggered magnetic moment and Néel temperature versus interchain coupling $|J_b|$, with J kept constant at 136.5 meV The solid curves represent the mean field expressions. The points correspond to Table III.

FIG. 4. Staggered magnetic susceptibility in system of weakly coupled AF chains. Shown are $\chi_{st}^{1D}$, $\chi_{st}^{2D,RPA}$ (Eq. 10) and an estimate of $\chi_{st}^{2D}$ that interpolates between $\chi_{st}^{2D,RPA}$, for $T>T^*$, and $1/T \exp(AJ/T)$ with $A\sim0.25$ and $J=\sqrt{J_aJ_b}$, for $T<T^*$.

FIG. 5. Néel temperature $T_N$ as a function of $J_c$ for $J_b=0.1$ (all in units of $J_a$) as predicted by Schulz's theory and by Linear Spin Wave theory. The intersection of the two curves determines the LSW prediction of $J_c^*$.

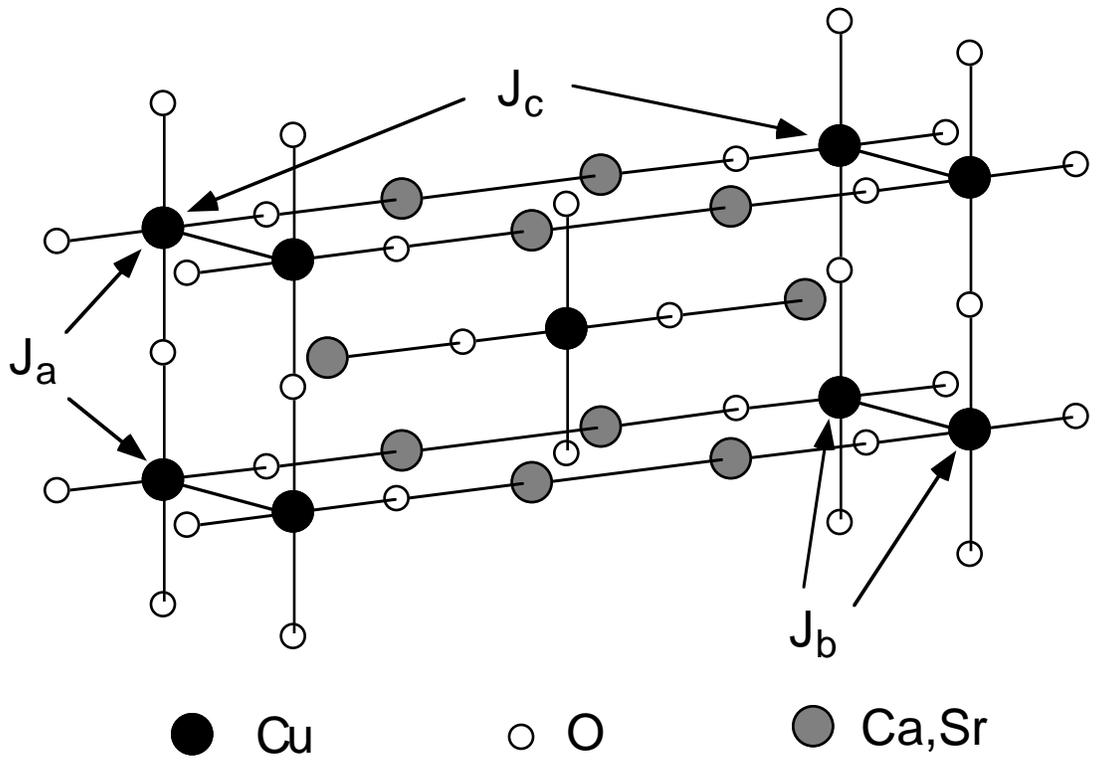

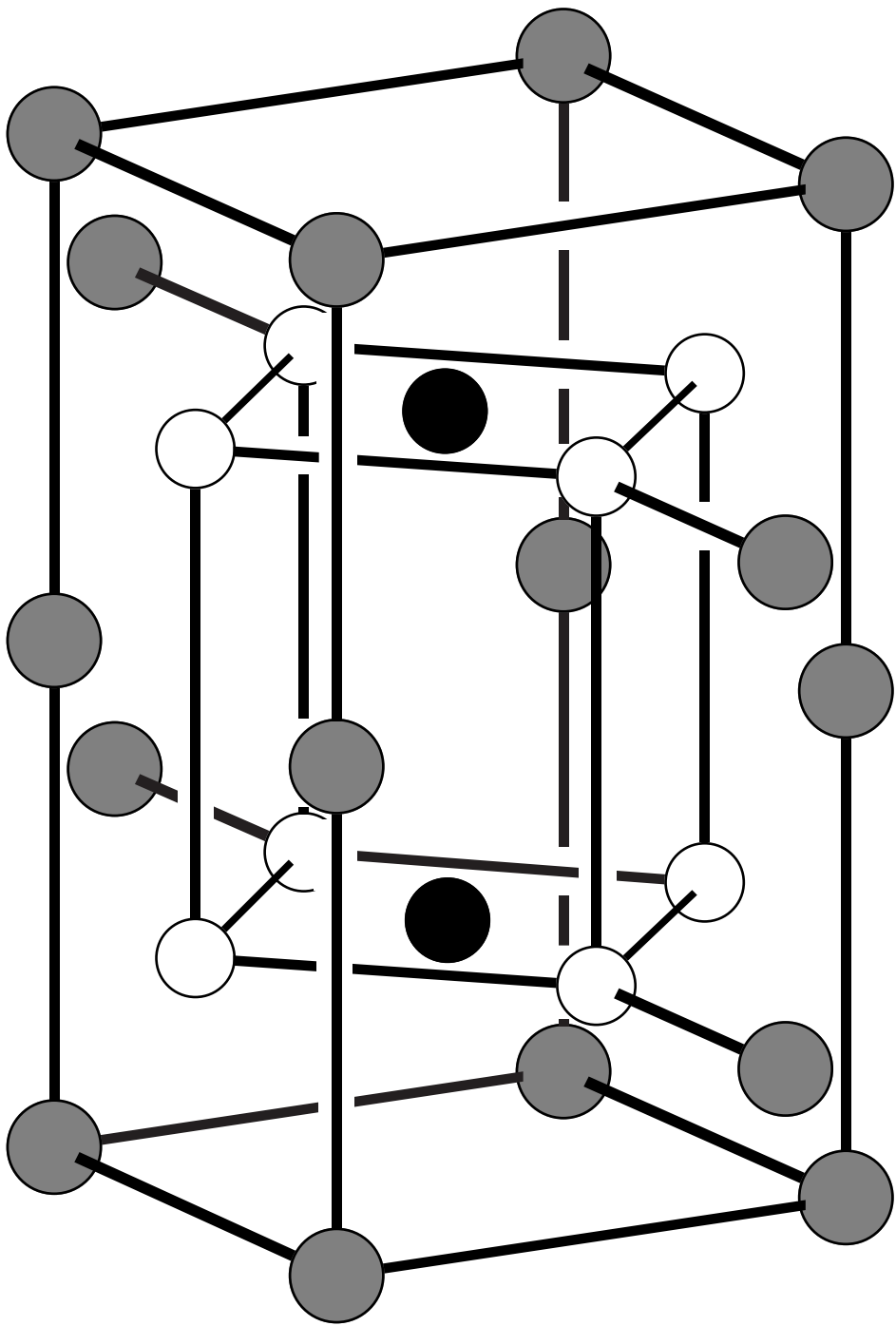

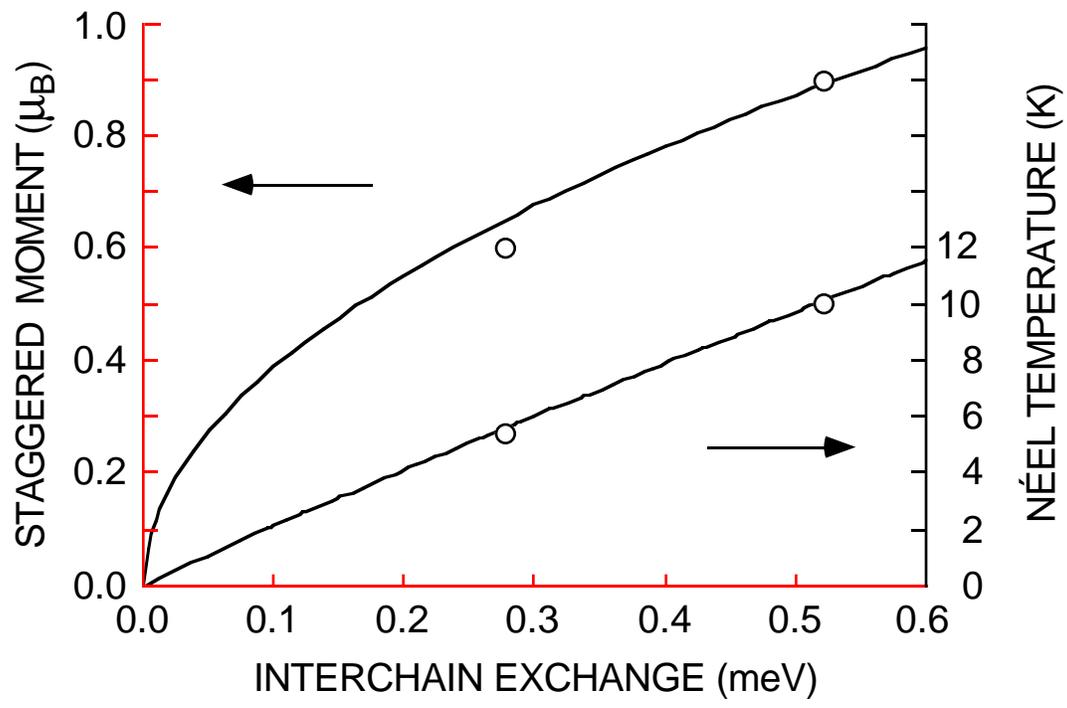

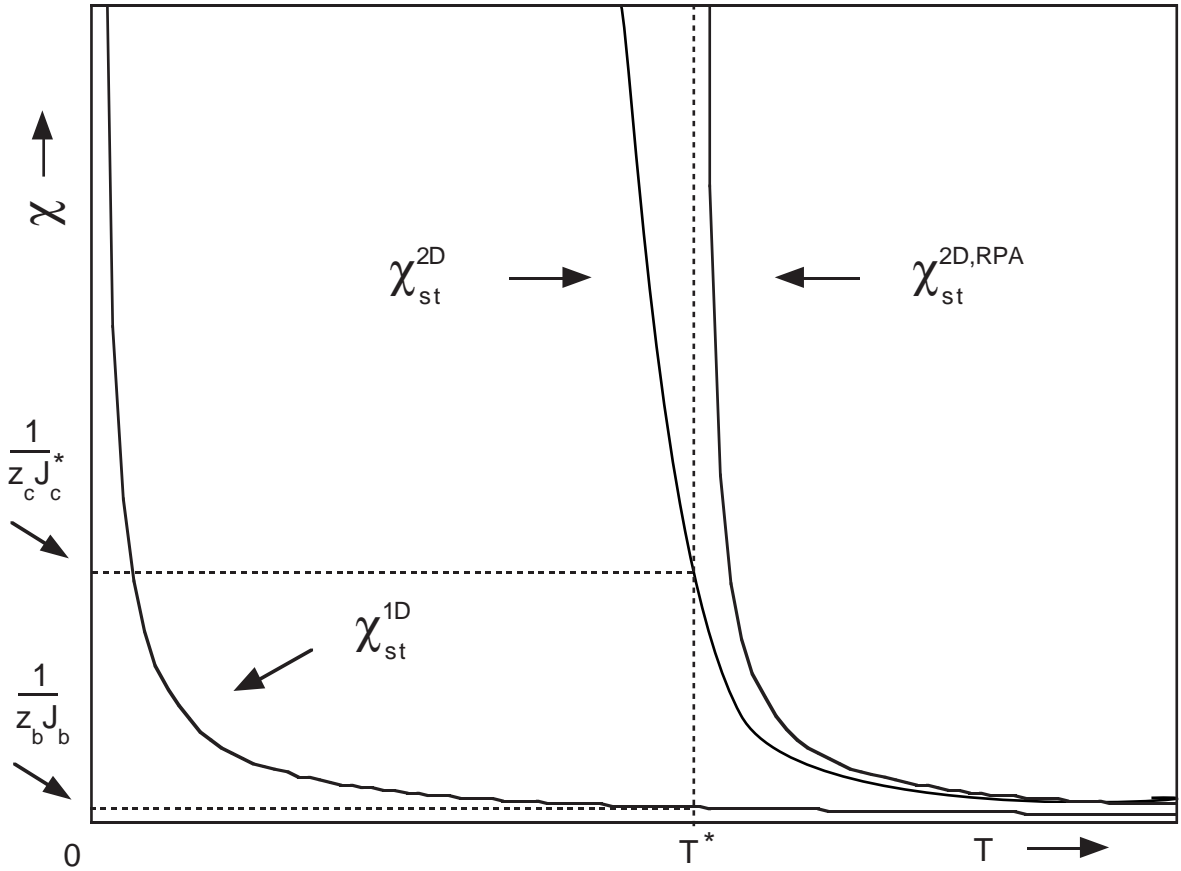

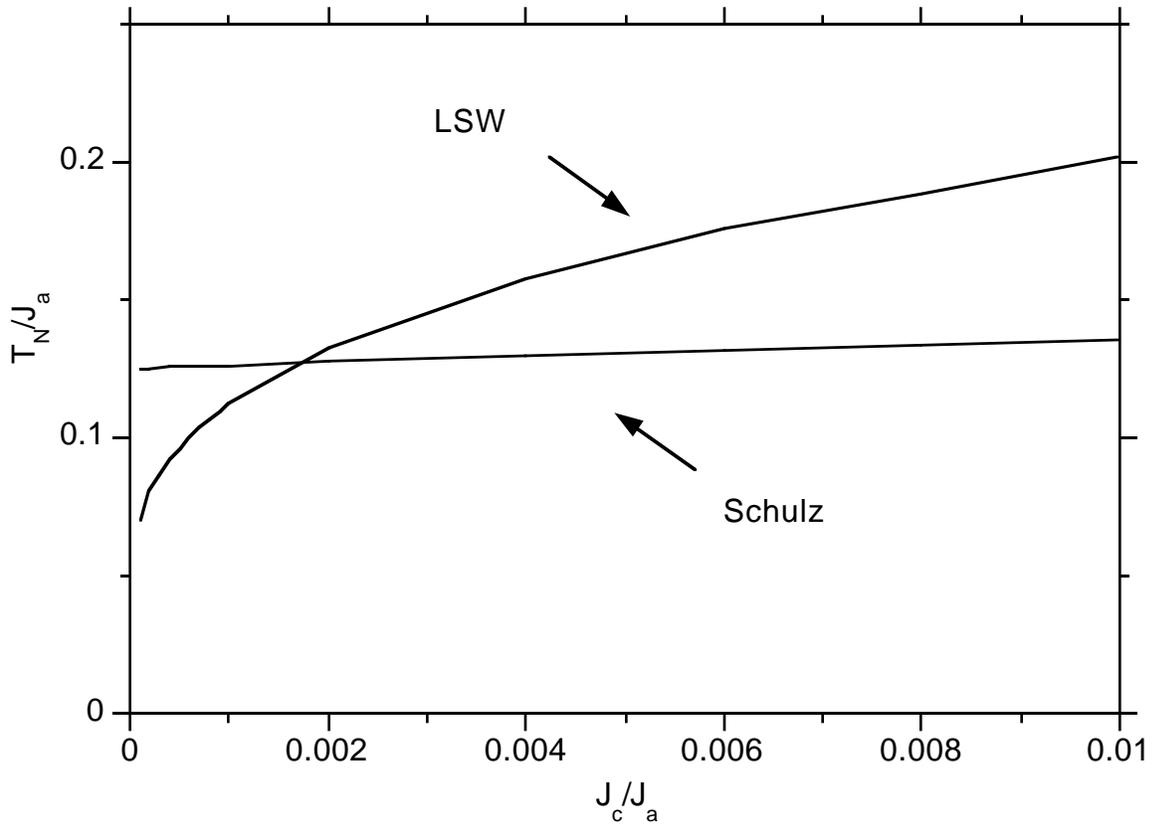